# Volumetric-mapping-based inverse design of 3D architected materials and mobility control by topology reconstruction

Kai Xiao[1], Xiang Zhou[2]*, Jaehyung Ju[1]*

1. UM-SJTU Joint Institute, Shanghai Jiao Tong University, 800 Dongchuan Road, Shanghai, China

2. School of Aeronautics and Astronautics, Shanghai Jiao Tong University, 800 Dongchuan Road, Shanghai, China

**Abstract**: The recent development of modular origami structures has ushered in a new era for active metamaterials with multiple degrees of freedom (multi-DOF). Notably, no systematic inverse design approach for volumetric modular origami structures has been reported. Moreover, very few topologies of modular origami have been studied for the design of active metamaterials with multi-DOF. Herein, we develop an inverse design method and reconfigurable algorithm for constructing 3D active architected structures — we synthesize modular origami structures that can be volumetrically mapped to a target 3D shape. We can control the reconfigurability by reconstructing the topology of the architected structures. Our inverse design based on volumetric mapping with mobility control by topology reconstruction can be used to construct architected metamaterials with any 3D complex shape that are also transformable with multi-DOF. Our work opens a new path toward 3D reconfigurable structures based on volumetric inverse design. This work is significant for the design of 3D active metamaterials and 3D morphing devices for automotive, aerospace, and biomedical engineering applications.

## Introduction

With the advancement of conventional manufacturing in both additive [1] and subtractive [2] ways, the fabrication of complex 3D structures is being realized on nano [3-7], micro [8-12], meso [13-19], and large scales [20-22]. Researchers are even pushing the limit of complex 3D structural design to motion structures that can change their shapes from one to another states, tuning their physical properties in adapted and active ways to varying physical environments [23].

Many groups have employed origami and kirigami techniques, the ancient folding and cutting arts [24], given the potential of future intelligent reconfigurable structures. Starting from 2D flexible structures using a Miura-ori sheet [25], Waterbomb base [26], or Ron-Resch pattern [27], researchers have recently explored the design of 3D architected structures using modular origami. Stacking multiple origami sheets [28,29] and assembling foldable modules [15-16, 30-33] are the currently available methods to construct

3D architected structures. Stacked Miura-ori [28] and Tachi–Miura polyhedron [30] are deployable with flat-foldability while a single origami sheet serves as a building block. Assembled structures with folding modules using prismatic polyhedrons [15-16,33], tubular bellows [31,34-35], and kirigami [32] have shown the potential of isotropic design by spatial symmetry while displaying single and multiple degrees of freedom (DOFs) [15-16,33].

All these methods require the construction of a building block (unit cell) with crease patterns and spatial periodicity (tessellation). This bottom-up design approach is convenient for controlling the kinematic and kinetic properties of the overall 3D architected structures with only a unit-cell design. However, this bottom-up design approach has a critical limitation when constructing practical engineering and artistic structures, whose shapes are mostly 3D curvilinear (e.g., automotive and aerospace structures with various curvatures), which require that the size and shape of the building blocks no longer be homogeneous in the design domain. Notably, most rectangular or cubic engineering structures have curved edges for minimum stress concentration and safety. Because of the break in spatial periodicity in curvilinear geometries, the kinematic and kinetic properties of the unit cell tend to differ from the properties of the tessellated structure, limiting the structural implementation of the architected origami materials.

Very few studies have explored a top-down approach to the design of 3D architected origami structures [36]. The existing inverse-design methods can only be applied to 2D curvilinear surfaces with origami and kirigami [37-44] and not to volumetric 3D spatial geometries. In the current work, we explore a top-down approach for the design of architected origami materials. Without tessellating a constant building block, our inverse design produces volumetric gradient cells mapped into complex curvilinear 3D geometries. Furthermore, we can provide mobility to the architected structures built in the curvilinear 3D space, yielding reconfigurable architected structures. Unlike the traditional bottom-up method, our top-down approach can generate volumetrically gradient cells for spheres, cones, twisted cylinders, toruses, and any combined curvilinear shapes with tunable mobility. Even though we focus on the non-periodic structural application of architected materials in this work, our method can be universally applied to conventional periodic structural design.

Our inverse design can be used for various 3D active morphing structures in automotive, aerospace, and biomedical engineering applications and 3D active metamaterials to tune both the dynamic properties (acoustic and elastic wave) and static properties (directional stiffness).

**Inverse design of nonperiodic structures**
Figure 1 shows the procedure for the inverse design of architected materials with modular origami. Using a space-filling tessellation, we construct a reference template consisting of polyhedrons, as shown in Figure 1a. The reference template can consist of a single polyhedron or multiple polyhedrons. Next, we transform the reference template into a prescribed curvilinear geometry in Figure 1b, building a transformed template in Figure 1c. To construct the transformed template, we implement the optimum transport algorithm [45], obtaining deformed polyhedrons based on the minimum energy for deformation. Notably, the transformed template still has single or multiple combinations of deformed polyhedrons.

Now, we spatially shrink the deformed polyhedrons of the transformed template while implementing a scaling constraint, as illustrated in Figure 1d:

$$S_{a_{1,p}} = S_{b_{1,p}} = S_{a_{2,p}} = S_{b_{2,p}}, \tag{1}$$

with the scaling ratios $S_{a_{i,p}} = \frac{|a_{i,p} - A_p|}{|o_i - A_p|}$ and $S_{b_{i,p}} = \frac{|b_{i,p} - B_p|}{|o_i - B_p|}$ ($i = 1,2$). $o_i$ is the centroid of the deformed $i$-th polyhedrons in the transformed template. $A_p$ and $B_p$ belong to the edge of the $p$-th face shared by two adjacent polyhedrons in the transformed template. The scaling constraint in Equation (1) forces the direction of connection on the $p$-th face to be parallel after shrinking; i.e., $a_{1,p} - a_{2,p} = b_{1,p} - b_{2,p}$, which is the critical condition for building a connection with adjacent polyhedron units. Our work is different from previous work [16], where the connecting direction was normal to the regular polyhedron; instead, we determine the connecting direction by bridging the centroids of the transformed templates.

We can also obtain the length of the connection $L_p$ between the adjacent polyhedron units after shrinking:

$$L_p = S_{a_{1,p}} |o_2 - o_1|. \tag{2}$$

Next, we connect the shrunken polyhedrons by extruding prismatic tubes on each shared face inside the transformed template, as shown in Figure 1e. For the faces of the shrunken polyhedrons that initially lie on the boundary of the transformed template, for example, the boundary face indicated by nodes $b_{2,p}, c_{2,m}$ and $d_{2,m}$, any extruding direction can be applied as long as there is no intersection of prismatic tubes, and the extrusion distance usually matches the boundary of the template. Finally, we generate a modular origami structure with a curvilinear shape.

The connected tubes serve as structural components, whereas the shrunken polyhedrons function as porous holes. The inverse-design method of the spatial architected materials can also be applied to planar cases, as illustrated in Figure 1g. Notably, our design method is universal, meaning that any combination of polyhedrons can be used to build a reference template whether the target geometry is a curvilinear 3D shape or planar shape.

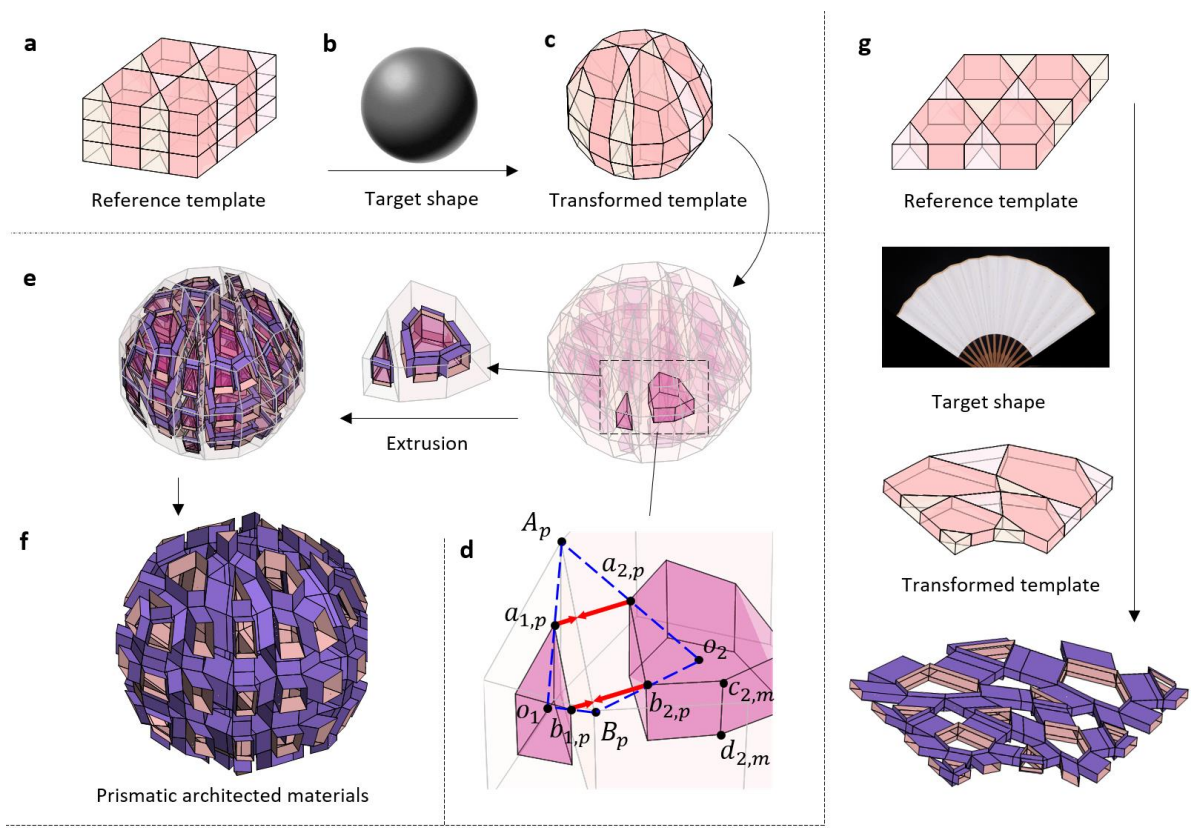

*Figure 1. Procedure of the inverse design of architected materials with modular origami: (a) reference template with polyhedrons, (b) a target shape, (c) transformed template with deformed polyhedrons. The details of volumetric mapping from 'a' to 'c' are provided in Section 1.1 "Optimal mass transport" of the Supplementary Information. (d) Shrunken polyhedron units highlighted in purple, (e) extrusion of prismatic tubes to connect adjacent polyhedron units, (f) an architected material of a sphere with modular origami, and (g) inverse design of a planar architected material with a different target shape.*

We can build complex 3D architected materials using the inverse-design algorithm based on various spatial-filling tessellations for volumetric mapping into an arbitrary target shape, as illustrated in Figure 2. Figure 2a demonstrates the volumetric mapping of the reference template with a single polyhedron set (cube) to generate a spherical structure; additional examples of the inverse design with the volumetric mapping are provided in Supplementary Fig. 5. Notably, the initial polyhedron in the reference template is involved in the nonhomogeneous deformation during the construction of the transformed template. Moreover, we can apply the volumetric mapping to generate a concave template, as illustrated in Figure 2b–2e.

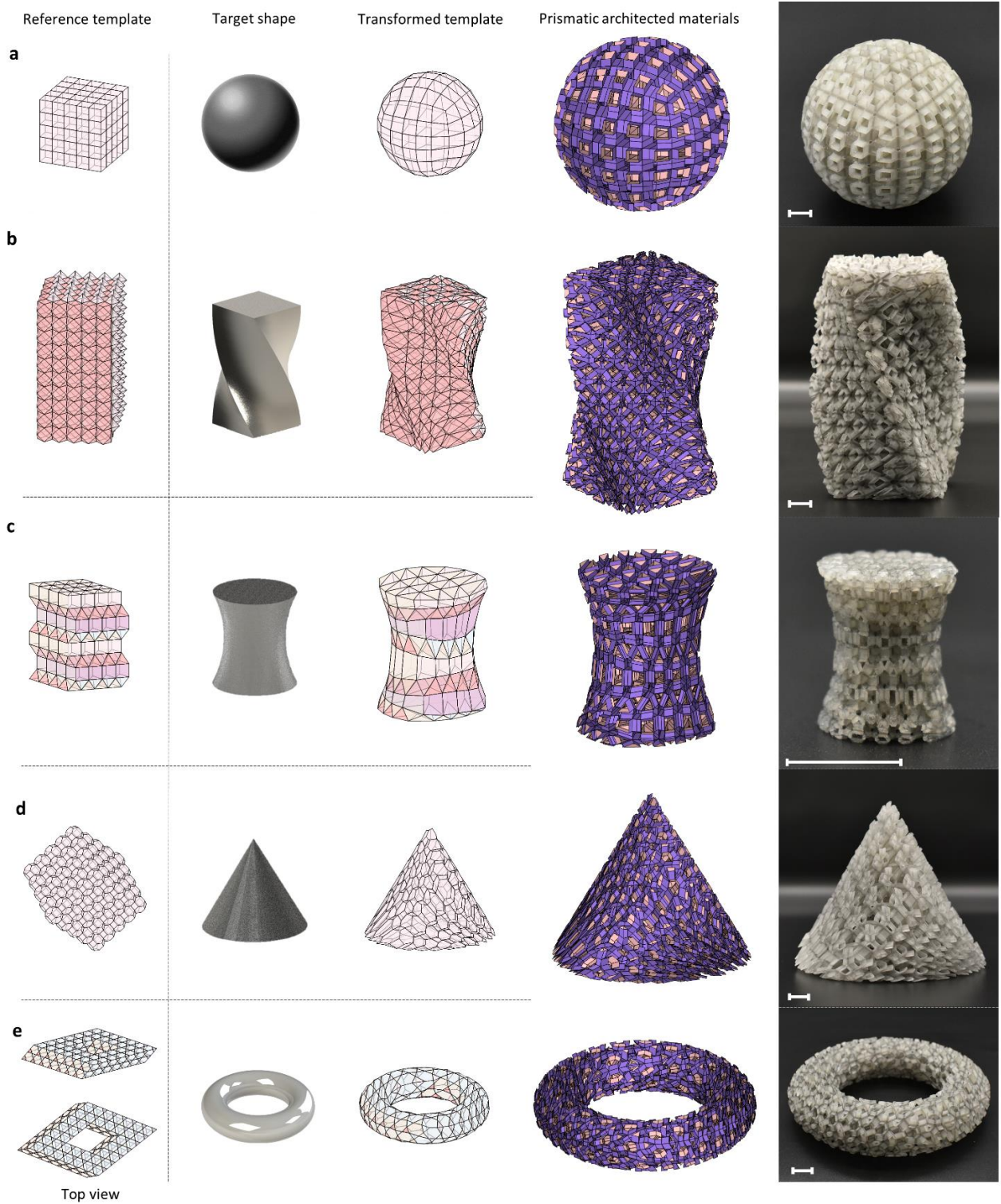

*Figure 2. Examples of inverse design of 3D architected materials with modular origami. The rightmost column shows the 3D-printed prototypes (See Section 1.2 "Fabrication of physical model" in the Supplementary Information). Scale bar, 1 cm.*

## Construction of tunable motion

Although our inverse-design method generates arbitrary 3D curvilinear structures through volumetric mapping, it hardly produces reconfigurability. From a macroscopic perspective, the fully connected origami modules (i.e., the structural unit) produce spatial loops and constrain inter-and intra-modular motions, delivering immobility to motion structures [46]. From a microscopic perspective, the irregular polyhedron units in the transformed template result in limited mobility (see Supplementary Figure 2), which is analogous to a kinematic linkage with lengths of different links and limited mobility according to Grashof theory [47].

To obtain 3D curvilinear structures with tunable motion, we developed a topology reconstruction method using a simple numerical algorithm. Adapting graph theory, we re-design the spatial loops of 3D curvilinear structures where neighboring modules are tightly connected. A graph consists of vertices and edges representing polyhedron units and tubular connections, respectively, among adjacent units. We selectively remove loops (cycles) from the graph at the connection of modules to provide mobility. We formulate the construction of the graph using a constrained optimization problem, solving it using a genetic algorithm tool in MATLAB; the objective of the algorithm is to identify graphs with a smaller number of basic cycles ($n_{bc}$) while retaining the structural integrity.

After screening the spatial loops, we reconstruct the microscopic topology of origami modules by selecting flexible ones from a library for each polyhedron unit of the transformed template. Note that the previous screening of spatial loops using graph theory does not cover a secondary connection by the geometry of modules. Combining the spatial loops with the selection of modules, we can generate 3D curvilinear structures with tunable motion (see details in Section 1.3, "The algorithm of enhancing reconfigurability" in the Supplementary Information.)

Figure 3a illustrates the procedure to construct a reconfigurable structure. A $2 \times 2 \times 2$ transformed template with a sphere comprises irregular tetrahedra and octahedra. We construct a fully connected graph for a sphere with $n_{DOF} = 0$. Then, we reconstruct the nodes with fewer loops and select modules with nonzero mobility in Figure 3b, producing a deformable assembly with three DOFs ($n_{DOF} = 3$), as shown in Figure 3c and Supplementary Video 2. As demonstrated by the numerical simulation and experiments in Figure 3d, this new structure can change its shape from a sphere to other shapes via rotational motion at the hinges (see the transformation process in Supplementary Video 3).

Figures 3e and 3f show the topology reconstruction procedure of transformed templates for other target shapes — a cone and hyperboloid. Notably, the cone and hyperboloid structures in Figures 3e and 3f have the same spatial connection topology and modules as the spherical motion structures in Figure 3d, producing the same $n_{DOF}$ and similar tunable motion (see Supplementary Video 4). Therefore, the tunable motion is controlled by the spatial connection and modules, not by the geometry of the transformed template.

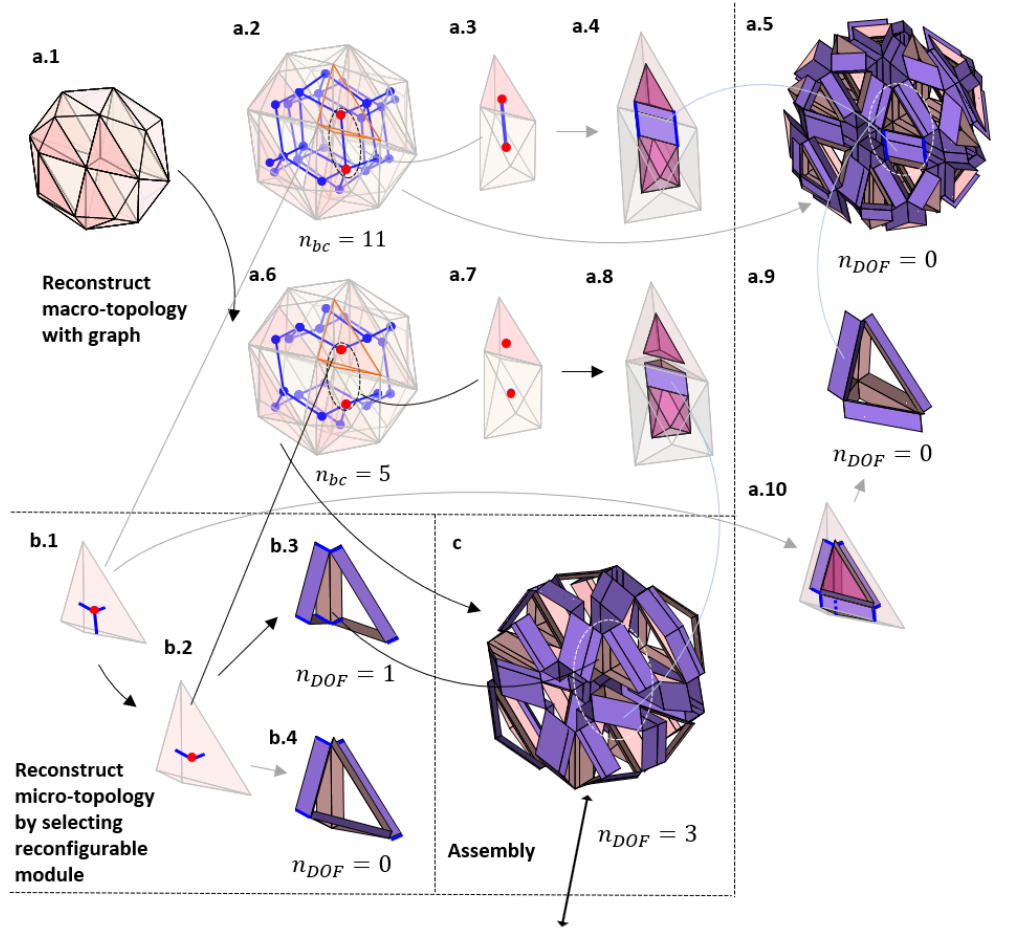

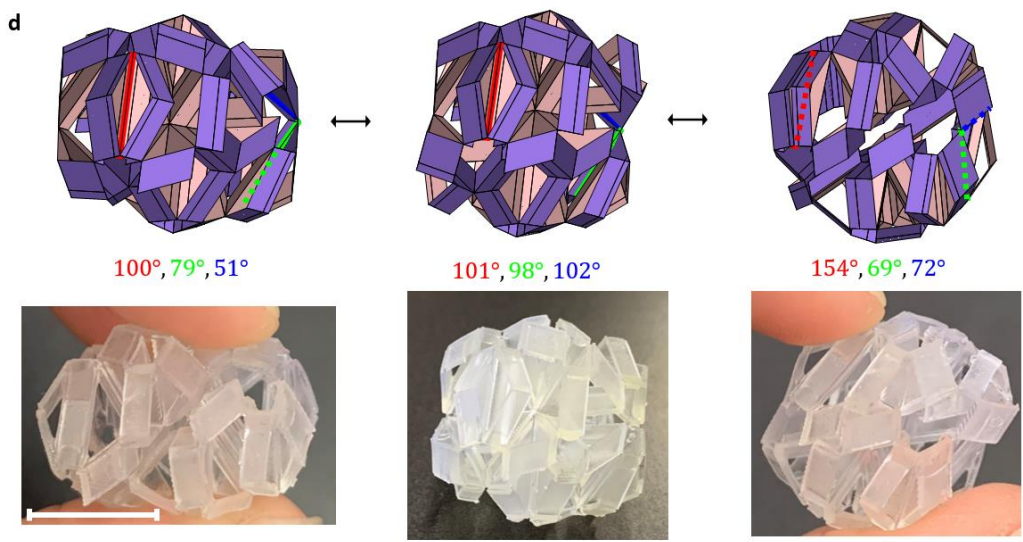

$100°, 79°, 51°$     $101°, 98°, 102°$     $154°, 69°, 72°$

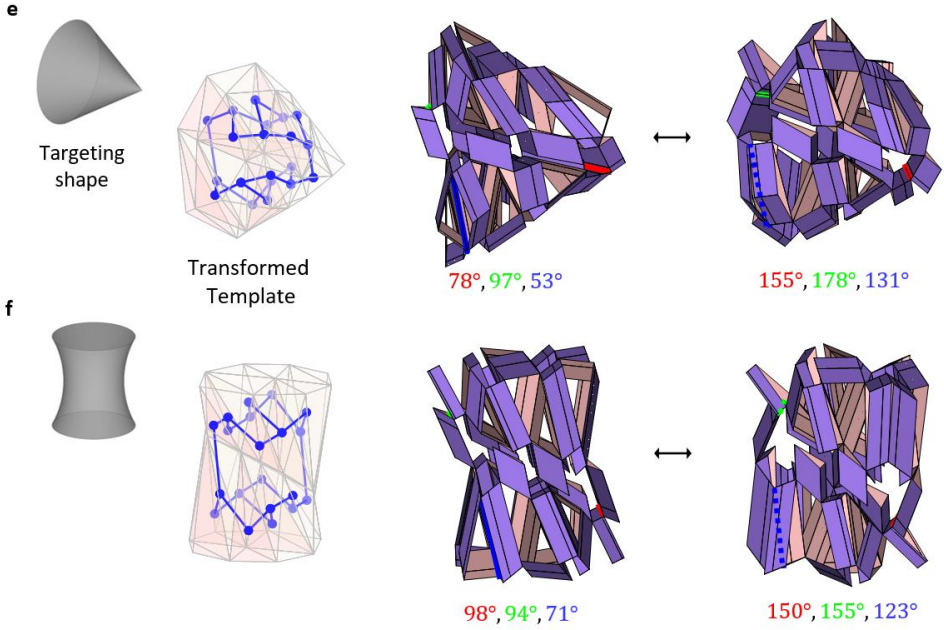

*Figure 3. Procedure to produce reconfigurability from transformed templates: (a.1) the transformed template in a spherical shape comprises irregular tetrahedra and octahedra; (a.2) construction of a fully connected graph with basic cycles: the number of basic cycles $n_{bc} = 11$; (a.3) an edge passing their symbolic nodes on two adjacent polyhedra for the graph; (a.4) faces of two adjacent shrunken polyhedrons and extrusion of tubes to form a connection; (a.5) the resulting assembly following the fully connected graph has zero DOFs; (a.6) a reconstructed graph with fewer basic cycles, where $n_{bc} = 5$; (a.7) two disconnected nodes; (a.8) two adjacent shrunken polyhedrons are disconnected; (a.9) a rigid module; (a.10) extruding process of the rigid module; (b.1) a polyhedron unit selected from the fully connected graph; (b.2) the same polyhedron unit from the graph with reduced $n_{bc}$; (b.3) selection of a module with nonzero mobility from a library; (b.4) a disqualified rigid module with the same connectivity as (b.3); (c) a deformable assembly with 3 DOFs; (d) the transformation of the structure c where the DOFs are denoted by three dihedral angles with colored hinges. See details of the physical prototypes in Section 1.2 "Fabrication of physical model" in the Supplementary Information. (e–f) Templates in cone and hyperboloid shapes having the same graph as a.7 with basic cycle, where $n_{bc} = 5$, producing a reconfigurable structure with the same DOFs ($n_{DOF} = 3$). Scale bar in d, 1 cm.*

We implement a numerical algorithm to identify mobility as a form of independent dihedral angles to simulate the transformation within the configuration space. Following the assumption of the rigid plate and flexible hinges of the prismatic modular structures, we identify the independent angles from a linearized constraint matrix. We apply kinematic constraints to the structures during transformation: the distance between two vertices on the edge remains constant, e.g., $|\mathbf{v}_1 - \mathbf{v}_2| = const$. The position vectors $\mathbf{v}_1, \mathbf{v}_2, \mathbf{v}_3$, and $\mathbf{v}_4$ on the surface remain in a plane, e.g., $(\mathbf{v}_1 - \mathbf{v}_2) \times (\mathbf{v}_3 - \mathbf{v}_2) \cdot (\mathbf{v}_4 - \mathbf{v}_1) = \mathbf{0}$ (see Supplementary Figure 4). Imposing the constraints on every vertex and linearizing them with a constraint matrix form $\mathbf{J}_v$, we obtain $\mathbf{J}_v \cdot d\mathbf{v} = \mathbf{0}$ [48], where $d\mathbf{v}$ is the infinitesimal displacement of vertices. We can also obtain the infinitesimal displacement of all the dihedral angles as $d\boldsymbol{\phi} = \mathbf{J}_h \cdot d\mathbf{v}$. In matrix form, the constraints can be expressed as

$$\begin{bmatrix} \mathbf{J}_v & \mathbf{0} \\ \mathbf{J}_h & -\mathbf{1} \end{bmatrix} \begin{bmatrix} d\mathbf{v} \\ d\boldsymbol{\phi} \end{bmatrix} = \mathbf{0}. \tag{3}$$

By calculating the reduced row echelon form of $\mathbf{J}\left(=\begin{bmatrix} \mathbf{J}_v & \mathbf{0} \\ \mathbf{J}_h & -\mathbf{1} \end{bmatrix}\right)$, we can determine the free variables in terms of $d\boldsymbol{\phi}$, which can produce the motion of the structures. Section 1.4 "Kinematics" of the Supplementary Information provides further details.

To investigate the tunable range of mobility of 3D prismatic architected materials, we applied our algorithm to transformed templates with 28 uniform space-filling tessellations [49,50]; we limited the template to a maximum of 40 polyhedra. We searched 360 reconfigurable structures in each transformed template by constructing graphs with a specific range of $n_{bc}$ ($2 \leq n_{bc} \leq 12$) and assembling different modules. The algorithm generated reconfigurability of a sphere motion structure for different transformed templates. For example, Figure 4 shows the reconfigurable structures constructed using the transformed templates #14 and #22, leading to DOFs ranging from 1 to 47 with varying deformation modes including shear and biaxial expansion (see other deformation modes of these examples in Supplementary Video 5). Our algorithm can be used to find the range of mobility of the motion structures by controlling the number of basic cycles $n_{bc}$ of the graph, as shown in Figure 4.

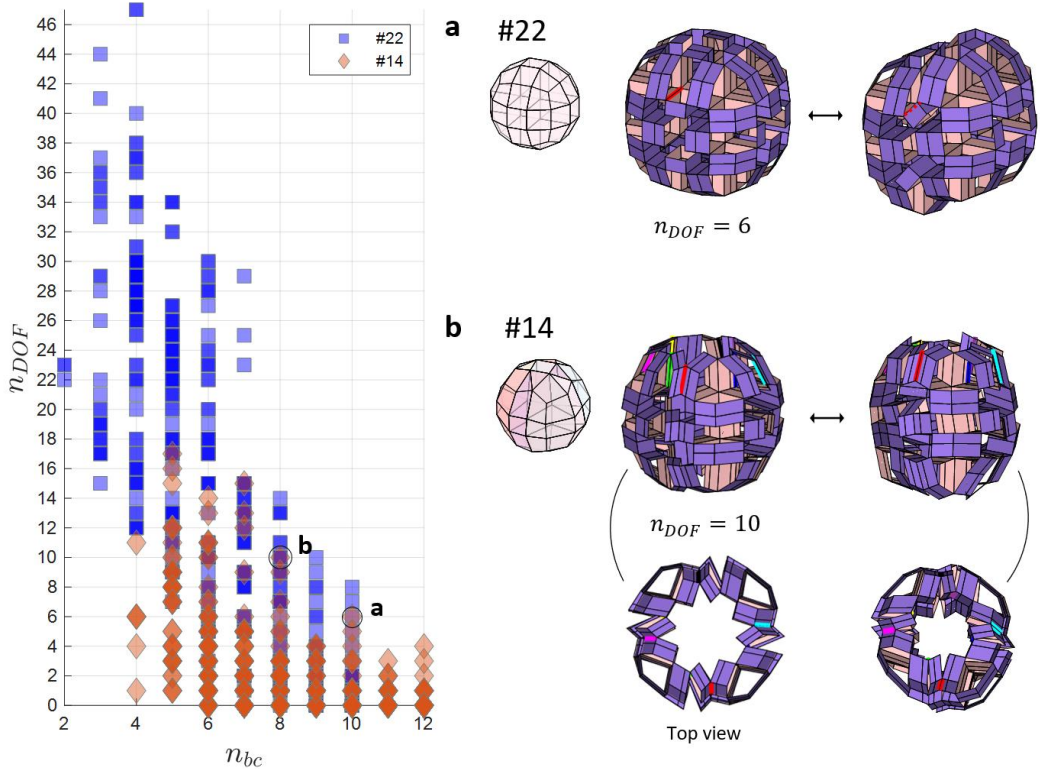

*Figure 4. Architected materials with diverse reconfigurability. Left-hand side: a scatter plot showing 400 reconfigurable designs for specific templates (#22 and #14). Each point represents one transformed template; the points with the same $n_{DOFs}$ and $n_{bc}$ overlap in the plot. Right-hand side: selective reconfigurable structures with (a) shear deformation and (b) biaxial expansion.*

## Discussion

3D curvilinear structures are ubiquitous in automotive, aerospace, and ocean engineering structures. To date, the design of architected materials has relied on the bottom-up design approach, using planar and orthogonal periodic tessellation of modules [23,51], which is convenient for structural design. However, this bottom-up approach has a critical limitation when constructing curvilinear 3D engineering and artistic structures, where the size and shape of the building blocks are no longer homogeneous in the design domain. The reconstruction of geometry and topology in this study provides a solution for the design of 3D modular origami for non-periodic and curvilinear 3D structures and their tunable motion. The geometric reconstruction by volumetric mapping of reference templates can be used to construct a stress-free 3D curvilinear structure. The fully connected irregular polyhedron units produce spatial loops and constrain inter-and intra-modular motions, delivering a rigid yet stiff curvilinear structure. The topology reconstruction of primary and secondary connections can yield tunable motion with multi-DOF. The tunable motion is controlled by the topology of the spatial connection and modules, not by the geometry of the transformed template.

The inverse design of 2D origami structures can generate a curvilinear surface with local control of the variable curvature [37,43]. Because of the instability (local snapping) of the out-of-plane deformation during reconfiguration, relatively high local energy is needed for the transformation [37]. The bottom-up approach with periodic modular origami structures can be used to build a curved shape by filling building blocks inside the curvilinear space [15,36]. However, this approach cannot yield a smooth surface; it only provides discrete curves, whereas the geometric reconstruction in this work can produce a smooth 3D curvilinear structure. Notably, the 3D curvilinear structures still cannot provide the exact 3D motions we desire at the initial design level because the topology reconstruction in this work is not fully integrated with the inverse design, which can be explored further.

The geometric and topology reconstruction can expand the design space of 3D modular origami, opening a new avenue for the design of 3D active metamaterials. The spatial scaling of space-filled polyhedrons can be applied for both orthogonal and curvilinear coordinates, meaning that our inverse design method is a universal approach that can be applied for both periodic and non-periodic structures. Our physical prototypes of 3D curvilinear structures fabricated by additive manufacturing validate the tunable motion, demonstrating untethered actuation by embedded hard magnets and moving external magnetic fields (see Supplementary Video 3 and 4). The noncontact actuation of motion structures demonstrates the potential of active metamaterials, morphing devices, and soft robots on the mesoscale. Advanced micro-fabrication techniques such as two-photon lithography [7] and micro-stereolithography [11] can be applied to our reconfiguration systems on the microscale, e.g., for the design of battery electrodes [52] and microelectronic mechanical systems [53].

## Conclusion

In summary, we introduce a top-down design approach of modular origami structures to produce 3D curvilinear structures with tunable motion. The volumetric-mapping-based inverse design can be used to construct stress-free 3D curvilinear structures. Topology reconstruction (primary connection) of spatial connection and selection of modules (secondary connection) in the library can yield tunable motion with multi-DOF. Our geometry and topology reconstruction method of 3D curvilinear motion structures can be directly applied for i) 3D active metamaterials to achieve tunable acoustic and static properties on the nano-, micro-, and mesoscale and ii) 3D morphing devices with curvilinear shapes in automotive, aerospace, and biomedical engineering applications. Our top-down approach can expand the design space of modular origami to 3D non-periodic structures and 3D curvilinear geometries — spheres, cones, twisted cylinders, toruses, overcoming the limitation of planar and spatial tessellations for periodic structural design. Unlike 2D reconfigurable curvilinear origami structures, our 3D motion structures can be reconfigured into other shapes with multi-DOF and are frustration-free. This work will greatly advance the design of 3D architected materials with reconfigurability, shaking up traditional periodic tessellation-based 3D metamaterial design.

## Acknowledgements

J.J. acknowledges the support received from the Shanghai NSF (Award # 17ZR1414700) and the Research Incentive Program of Recruited Non-Chinese Foreign Faculty by Shanghai Jiao Tong University. We thank Y. Du for computational design of the prismatic materials with prescribed thickness; we also thank M. Huang, J. Sun, Y. Du, Y. Yu, and H. Wang for their assistance with the fabrication of the 3D-printed prototypes.

## Author contributions

K.X., X.Z., and J.J. proposed and designed the research; X.Z. and J.J. supervised the project; K.X. and X.Z. designed the numerical calculations; X.Z. and J.J. designed the physical models and experiments; K.X.

performed the numerical calculations and experiments; K.X. and X.Z. wrote the initial draft; J.J. revised the manuscript; all the authors reviewed the manuscript.

# Supplementary Information

**Volumetric-mapping-based inverse design of 3D architected materials and mobility control by topology reconstruction**

Kai Xiao[1], Xiang Zhou[2]*, Jaehyung Ju[1]*

1. UM-SJTU Joint Institute, Shanghai Jiao Tong University, 800 Dongchuan Road, Shanghai, China

2. School of Aeronautics and Astronautics, Shanghai Jiao Tong University, 800 Dongchuan Road, Shanghai, China

1. Methods

    1.1. **Optimal mass transport**

To perform volumetric mapping between the reference template and target shapes, we used an open-source software [54] (Graphite ver. 3, developed by Bruno Lévy), which conducts semi-discrete optimal transport [45]. As the input of the software, we prepared 3D object files of the boundaries of both the reference template and target shape using a custom-made MATLAB script. Then, we imported and volumetrically meshed the object files in the software and aligned the size and position of both meshes using a pre-processing tool in Graphite. Next, we set the density for sampling the mesh of the reference template to 400000 and transported the sampled mesh to fit the target shape. The software outputs a transported sample, which records the positions of these 400000 nodes in both the initial geometry and target shape. However, our template usually has much fewer nodes; building the transformed template requires matching the nodes of the reference template with the sample in the initial shape, which is achieved using the *knnsearch* tool in MALTAB. By replacing the positions of the nodes of the reference template to their counterparts in the transported sample with the target shape, we obtained the transformed template.

## 1.2. Fabrication of physical model

We used two types of additive manufacturing techniques for prototyping: 1. multi-materials inkjet printing (MultiJet from 3D systems, ProJet MJP 5600) and 2. stereolithography (SLA) (formlabs, Form 3). To generate the input file for 3D printing, we wrote a MATLAB script that converts the thin-walled model (Supplementary Fig. 1a and 1e) into one with prescribed wall thickness (Supplementary Fig. 1b and 1f). For the MultiJet printing, the thicknesses of the plate $t_p$ and hinges $t_h$ can be close ($0.7 t_p \leq t_h \leq t_p$), as shown in Supplementary Fig. 1c, because we assigned a rigid polymer (VisiJet CR-CL with Young's modulus $E \approx 1.3$ GPa) for the plate and a flexible one (VisiJet CE-BK with $E \approx 0.3$ MPa) for the hinges of the architected materials. For SLA, we set $0.05\, t_p \leq t_h \leq 0.5\, t_p$ and assigned a single soft material (Flexible 80A with $E \approx 2.3$ MPa) for the entire model. The printed physical models obtained using both approaches are shown in Supplementary Fig. 1d and 1h.

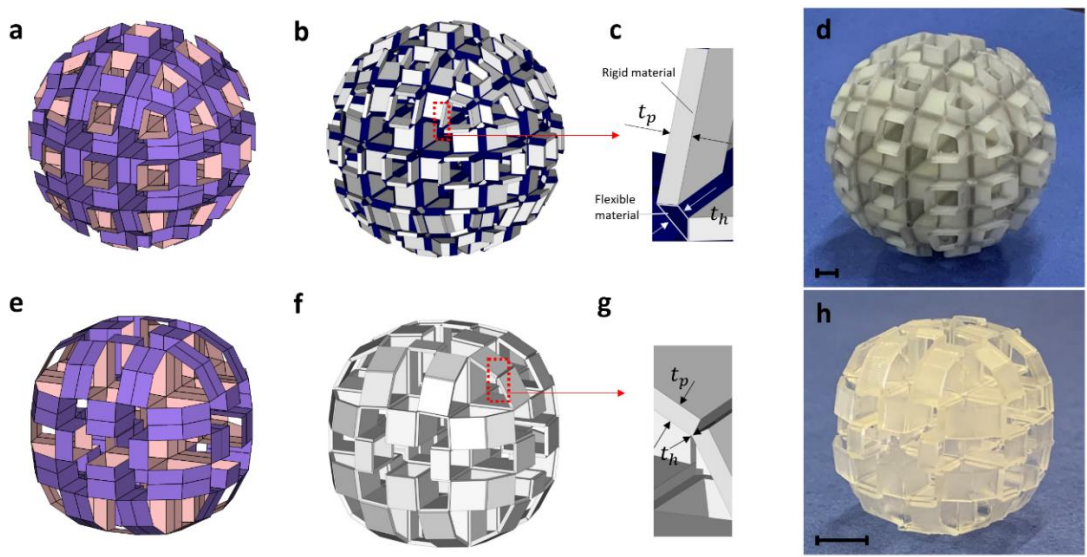

*Supplementary Fig. 3. (a) spherical prismatic architected material based on cubic tessellation. It is zero-thickness. (b) Mesh file of (a). (c) Schematics of $t_p$ and $t_g$. (d) Prototype based on MultiJet printing. (e) Spherical prismatic material with enhanced reconfigurability. (f) Mesh file of (e). (g) Schematics of $t_p$ and $t_g$. (h) Model based on SLA.*

### 1.3.  Algorithm to enhance reconfigurability

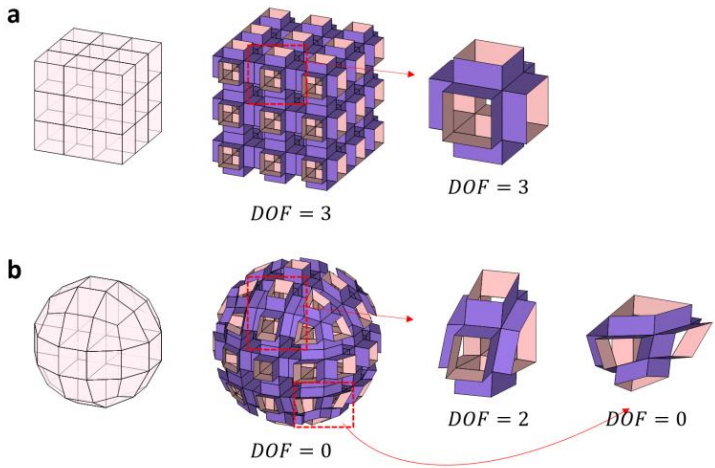

*Supplementary Fig. 2. Example of reduced mobility of the unit from the non-regular shape of the unit. (a) Cubic reference template, where each polyhedron has a regular and periodic shape, leading to assembled architected materials and unit with $DOF = 3$. (b) Transformed template, where each polyhedron has non-regular faces and a non-periodic shape, building the unit and assembly with reduced DOFs.*

**Algorithm to enhance mobility**

Supplementary Fig. 3 shows the flow of our algorithm to enhance the mobility of prismatic architected materials. As depicted in the figure, there are two inputs to our algorithm: the transformed template and the desired number of basic cycles ($n_{bc}$) for constructing the graph. The output is a reconfigurable prismatic material based on the resulting graph, where each origami module is reconfigurable. Between the inputs and output, there are three steps: 1. reconstruct a graph with a smaller number of basic cycles $n_{bc}$ to reduce the connections among units, 2. build a library consisting of reconfigurable modules at every node of the graph, and 3. assemble these modules without introducing extra cycles. The details of each step are discussed in the following sections.

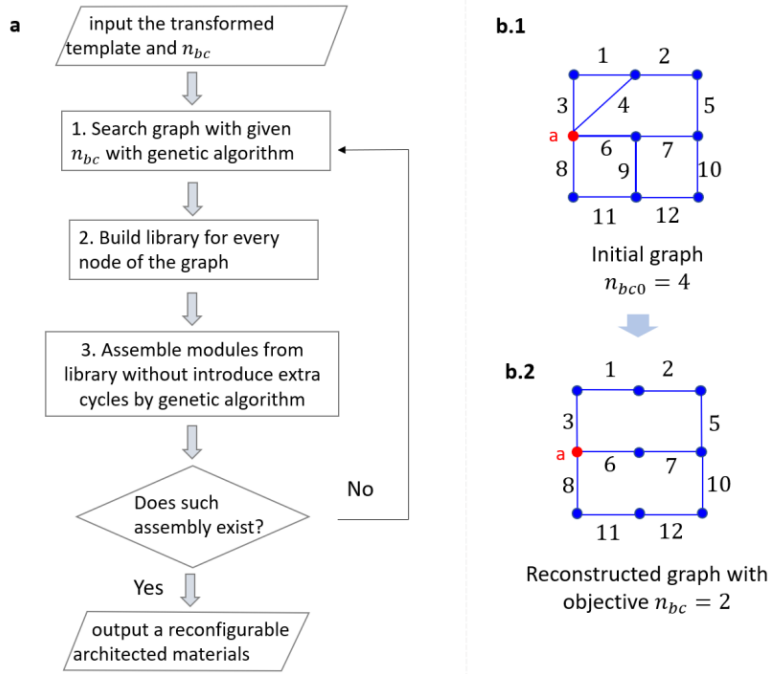

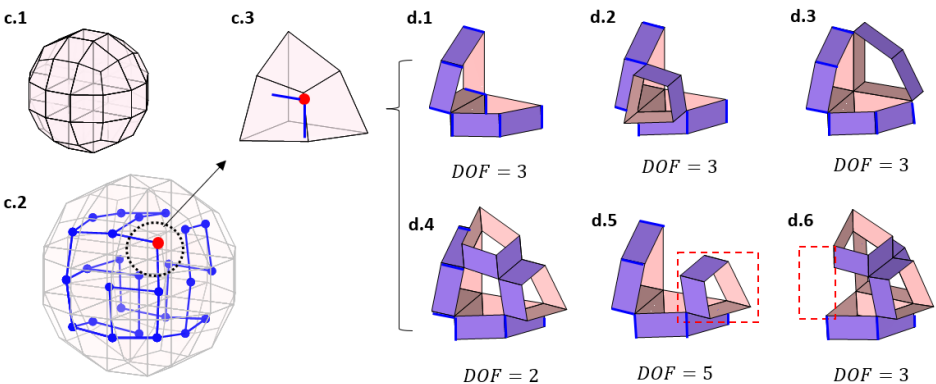

| Node $a$ | Module #1 | Module #2 | … | #A |
| --- | --- | --- | --- | --- |
| Node $b$ | #1 | #2 | … | #B |
| ⋮ | ⋮ | ⋮ | ⋮ | ⋮ |
| Node $N$ | #1 | #2 | … | #N |

*Supplementary Fig. 3. (a) Algorithm for finding a reconfigurable architected material with desired $n_{bc}$. (b) Schematics showing the reconstruction of a graph. (c.1) Transformed template, (c.2) its reconstructed graph, and (c.3) connectivity relationship on a node. (d) Flexible origami modules based on polyhedra unit in (c.3). (e) Chart illustrating the concept of selecting modules for final assembly.*

We input an initial graph with $n$ number of nodes and $e$ edges from the transformed template in procedure 1 and formulated the reconstruction of the graph as a constrained optimization problem. The objective of this optimization problem was to identify a graph with desired $n_{bc}$, which is usually smaller than the number of basic cycles $n_{bc0}$ of the initial graph. Then, we set the binary design parameter $x_1, x_2, x_3, \ldots, x_e \in \{0,1\}$ to represent the existence of edges in the initial graph during the reconstruction. Therefore, assigning different values of $x_1, x_2, x_3, \ldots, x_e$ generates different graphs; for the initial graph, $x_1, x_2, x_3, \ldots, x_e = 1$. Moreover, the current number of basic cycles $u_{bc}$ of the graph during reconstruction is a function of $x_1, x_2, x_3, \ldots, x_e$, and we write the optimization problem as

$$\min \quad |u_{bc}(x_1, x_2, x_3, \ldots, x_e) - n_{bc}| \tag{S1}$$

$$s.t. \quad \sum_{i}^{e} x_i > 0 \tag{S2}$$

$$2 \leq V_i \leq C_i \quad \text{for } i \in \{1, 2, \ldots, n\} \tag{S3}$$

$$N_{cc} - 1 = 0 \tag{S4}$$

$$N_{bic} = 0 \tag{S5}$$

$$N_{tri} = 0 \tag{S6}$$

The constraints shown in $S2$ to $S6$ are the structural requirements of the graph:

1. The constraint in $S2$ ensures the existence of the graph.
2. The inequalities $S3$ constrain the amount of connection on each node, where $V_i$ is the connectivity, i.e., the number of edges connected to the $i$-th node in the graph, and $C_i$ is the connectivity of the $i$-th node in the initial graph; we set the lower bound of $V_i$ to 2 to ensure that the $i$-th node is stringed by a cycle. The constraints in $S3$ only apply to the nodes we want to preserve in the reconstructed graph. In other words, we do not need to preserve all the nodes from the initial graph in the reconstructed graph. For the sake of completeness of the target shape, the nodes associated with the polyhedrons on the boundary of the transformed template are prior to preserve.
3. The constraint in Eq. $S4$ avoids identifying a disconnected graph. $N_{cc}$ is the number of connected graph components in the reconstructed graph, which can be calculated using the built-in function *conncomp* in MATLAB.
4. The constraint in Eq. $S5$ avoids the existence of biconnected components of a graph, where $N_{bic}$ is the number of biconnected components, which can be calculated using the built-in function *biconncomp* in MATLAB.
5. The constraint in Eq. $S6$ ensures that there are no triangular cycles consisting of only three nodes in the graph, as these triangular connections reduce the mobility.

As a simple example of reconstructing a graph, Supplementary Fig. 3b shows the difference between the initial graph and reconstructed graph. In this case, the number of edges in the reconstructed graph is reduced because the number of basic cycles $n_{bc}$ decreases to 2. Notably, the triangular cycle formed by edges 1, 3, and 4 is removed after reconstruction, and the two basic cycles in Supplementary Fig. 3b.2 are connected but not in biconnected form because of the structural requirement from constraints $S4$, $S5$, and $S6$. In addition, the connectivity of node $a$ is reduced to 3 from 4.

The next step after reconstructing the graph is to build a library consisting of mobile origami modules for every node of the graph. In each library, the desired modules are mobile (i.e., $n_{DOF} > 0$), satisfying the connectivity condition of the corresponding node, and have no solely connected prismatic tubes to avoid local motion. We find these modules by collecting possible partial extrusion configurations of the unit polyhedron. For example, Supplementary Fig. 3d shows the candidate modules for the library of the node (Fig. 3c.3) with two connections in the template in

Fig. 3c. Although all the modules shown in Supplementary Fig. 3d are flexible, two modules must be excluded. The one shown in Fig. 3d.5 has a solely connected prismatic tube, as highlighted by the red rectangle, leading to local DOFs for the reconfigurable system. In addition, the one in Fig. 3d.6 does not satisfy the connectivity. Therefore, the library of this node should include the modules shown in Fig. 3d.1–d.4.

We formulate the final procedure of assembling modules as another constrained optimization problem and solve it using the genetic algorithm tool *ga* in MATLAB. As indicated in Supplementary Fig. 3e, we want to select only one module from the library for each node and assemble them together without introducing extra cycles that are not indicated in the graph. To this end, we set the design parameters $y_1, y_2, \ldots, y_N$, which represent the identification index of modules for the nodes #1, #2, ..., #N in the graph. Specifically, for the $i$-th node with total $I_i$ number of modules in the library, its design parameter $y_i \in \{1,2,3,\ldots,I_i\}$. Furthermore, the objective is to minimize the number of extra cycles by varying $y_1, y_2, \ldots, y_N$, i.e.,

$$\min |N_{bc}(y_1, y_2, \ldots, y_N) - u_{bc}|. \tag{S7}$$

Here, $N_{bc}$ is the current number of basic cycles in the graph associated with the assembly, and $n_{bc}$ is the desired number of basic cycles. By doing so, we obtain the origami assembly having the same connecting topology as indicated in the graph.

### 1.4. Kinematics

**Analysis of DOFs and identifying independent angles**

To identify the DOFs of reconfigurable architected materials comprising rigid plates and hinges, we calculate the number of free variables associated with the linearized constraint matrix [48]. Specifically, we first triangulate each face of the structure to ensure the rigid plate constraints, for which an additional diagonal edge is generated on the face, as shown in Supplementary Fig. 4a. We then describe the length constraint on two vertices on the $i$-th edge as $(\mathbf{v}_{a_i} - \mathbf{v}_{b_i}) \cdot (\mathbf{v}_{a_i} - \mathbf{v}_{b_i}) = L_i^2$, for $i = 1,2,3,\ldots,E$, where $\mathbf{v}_{a_i}$ and $\mathbf{v}_{b_i}$ are the position vectors of the two vertices, $L_i$ is the length of the $i$-th edge, and $E$ is the number of total edges. Next, we prescribe the constraints arising from flat faces such that all the vertices on the $j$-th face remain planar during the transformation, i.e., $(\mathbf{v}_{4_j} - \mathbf{v}_{2_j}) \cdot \left[(\mathbf{v}_{1_j} - \mathbf{v}_{2_j}) \times (\mathbf{v}_{3_j} - \mathbf{v}_{2_j})\right] = 0$ for $j = 1,2,3,\ldots,F$, where $F$ is the total number of faces. After linearizing the above two types of constraints and assembling them into a matrix $\mathbf{J}_v$, we obtain $\mathbf{J}_v \cdot d\mathbf{v} = \mathbf{0}$, where $d\mathbf{v}$ is a vector that includes the infinitesimal displacement of every vertex. Finally, the number of DOFs is obtained as $3 \cdot n_v - \text{rank}(\mathbf{J}_v) - 6$, where $n_v$ is the total number of vertices in the structure, and the rank of $\mathbf{J}_v$ indicates the number of independent constraints.

The transformation process of a multi-DOF reconfigurable structure can be simulated by folding the dihedral angles of the independent hinges. To find such independent hinges, we combine $d\boldsymbol{\phi} = \mathbf{J}_h \cdot d\mathbf{v}$, where $d\boldsymbol{\phi}$ is the infinitesimal change in all the hinge angles, with the equation $\mathbf{J}_v \cdot d\mathbf{v} = \mathbf{0}$, leading to $\mathbf{J} \cdot \begin{bmatrix} d\mathbf{v} \\ d\boldsymbol{\phi} \end{bmatrix} = \mathbf{0}$. Here, we have $\mathbf{J} = \begin{bmatrix} \mathbf{J}_v & \mathbf{0} \\ \mathbf{J}_h & -\mathbf{1} \end{bmatrix}$, as mentioned in the main text. Specifically, for $d\boldsymbol{\phi} = \mathbf{J}_h \cdot d\mathbf{v}$, the Jacobian matrix $\mathbf{J}_h$ has the following entries:

$$J_{h[r, 3\cdot(n-1)+j]} = \frac{\partial \phi_r}{\partial v_{j,n}}, \tag{S8}$$

where $j = 1,2,3$. $dv_{j,n}$ is the displacement of the $n$-th vertex along the $j$-th axis of a Cartesian coordinate, $r = 1, \ldots T$. $T$ is the total number of selected hinges, and $n = 1,2,3 \ldots, N$. $N$ is the total number of vertices in the structure. A detailed derivation of $\mathbf{J}_h$ can be found elsewhere [16].

Next, we find the free column in the reduced row echelon form of $\mathbf{J}$. The free column corresponds to free variables in $\begin{bmatrix} d\mathbf{v} \\ d\boldsymbol{\phi} \end{bmatrix}$. In fact, all the free variables can be found in $d\boldsymbol{\phi}$, i.e., the independent angles.

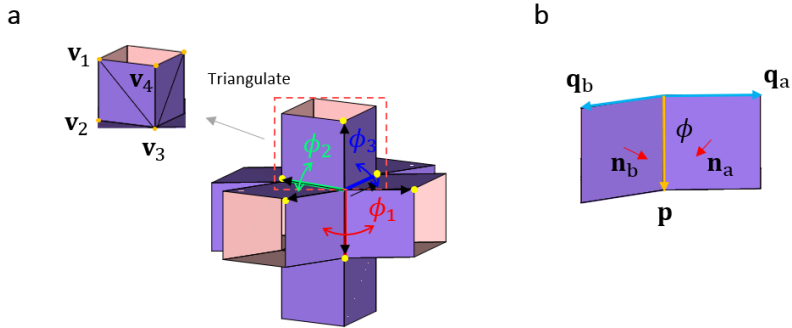

*Supplementary Fig. 4. (a) Schematic of triangulated face of a prismatic unit. (b) Dihedral angle with associated vectors.*

**Transformation simulation**

We obtain the transformed configurations of such reconfigurable architected materials using a numerical iterative method that applies a projection matrix to reduce numerical errors [55]. In each iteration, the input is a small increase $\begin{bmatrix} \mathbf{0} \\ d\boldsymbol{\phi}_0 \end{bmatrix}$ in the actuating angles, and the output is a displacement vector $\begin{bmatrix} d\mathbf{v} \\ d\boldsymbol{\phi} \end{bmatrix}$ satisfying the kinematics constraints. Between the input and output, there is

$$\begin{bmatrix} d\mathbf{v} \\ d\boldsymbol{\phi} \end{bmatrix} = -\mathbf{J}^+\mathbf{r} + [\mathbf{I} - \mathbf{J}^+\mathbf{J}]\begin{bmatrix} \mathbf{0} \\ d\boldsymbol{\phi}_0 \end{bmatrix}, \tag{S9}$$

where $[\mathbf{I} - \mathbf{J}^+\mathbf{J}]$ is the projection matrix, $\mathbf{J}^+$ is the pseudo-inverse of $\mathbf{J}$, and $\mathbf{I}$ is a unit matrix. The term $-\mathbf{J}^+\mathbf{r}$ in Eq. S9 is for reducing the numerical error in each iteration with a residual vector $\mathbf{r} =$

$[r_{e_1}, r_{e_2}, ..., r_{e_E}, r_{f_1}, r_{f_2}, ..., r_{f_F}]^T$, where $r_{e_i} = (\mathbf{v}_{a_i} - \mathbf{v}_{b_i}) \cdot (\mathbf{v}_{a_i} - \mathbf{v}_{b_i}) - L_i^2$ is the residual of the constraint of the $i$-th edge, and $r_{f_j} = (\mathbf{v}_{a_j} - \mathbf{v}_{2_j}) \cdot [(\mathbf{v}_{1_j} - \mathbf{v}_{2_j}) \times (\mathbf{v}_{3_j} - \mathbf{v}_{2_j})]$ is the residual of the constraint for the $j$-th face. $\mathbf{r} = [\mathbf{0}]$ in the first iteration.

**Non-overlapping requirement and configuration space**

To avoid overlapping of the faces during the transformation, we limit the range of dihedral angles of every hinge to $0 \leq \phi \leq \pi$, where

$$\phi = \pi - \tan^{-1}\left(\mathbf{p} \cdot \frac{(\mathbf{n}_a \times \mathbf{n}_b)}{\mathbf{n}_a \cdot \mathbf{n}_b}\right). \tag{S10}$$

Here, the face normals $\mathbf{n}_a = \frac{\mathbf{p} \times \mathbf{q}_b}{|\mathbf{p}| \cdot |\mathbf{q}_b|}$ and $\mathbf{n}_b = \frac{\mathbf{q}_b \times \mathbf{p}}{|\mathbf{q}_b| \cdot |\mathbf{p}|}$, as illustrated in Supplementary Fig. 4b. While inputting the increase of actuating angles in Eq. S9 leads to a transformed shape of the architected materials, the inequality $0 \leq \phi \leq \pi$ must be checked for every hinge angle to determine if the non-overlapping condition is satisfied. Finally, these valid transformations form the configuration space.

## 2. Supplementary results

**Prismatic architected structures based on 28 3D space-filling tessellations with spherical target shape**

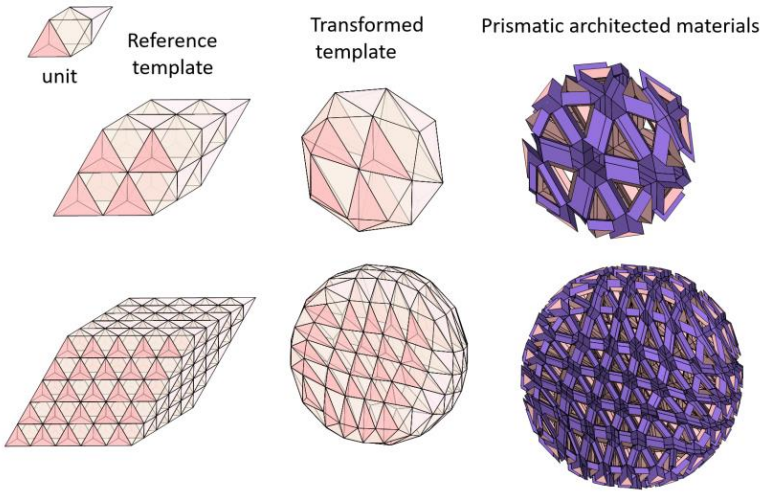

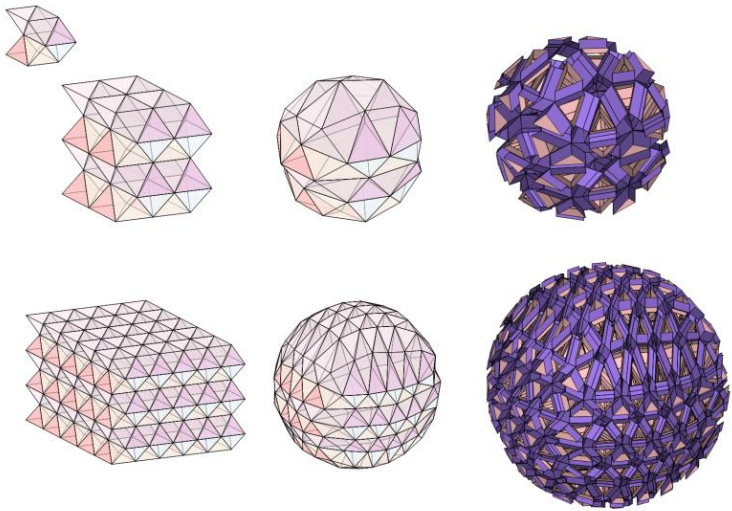

Supplementary Fig. 5: Inverse design of architected materials based on deforming 28 uniform space-filling tessellation template with a spherical shape.

#3 tetrahedra and octahedra and triangular prisms (2:1:2)

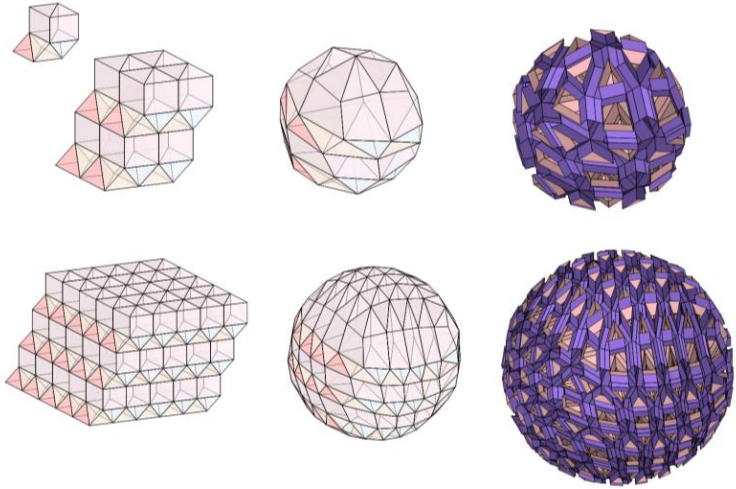

#4 tetrahedra and octahedra and triangular prisms (4:2:4)

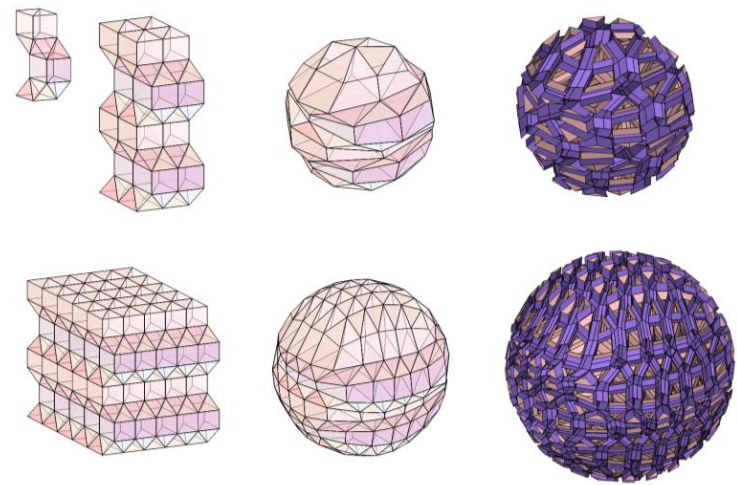

#5 tetrahedra, rhombicuboctahedra and cubes (2:1:1)

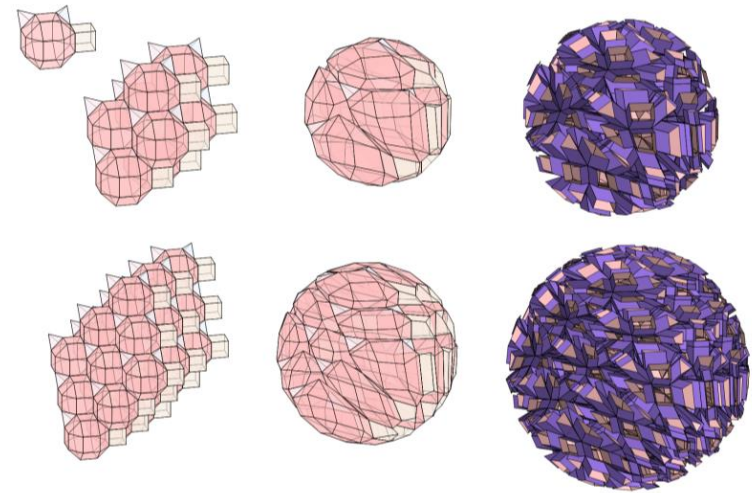

#6 tetrahedra and truncated tetrahedra (2:2)

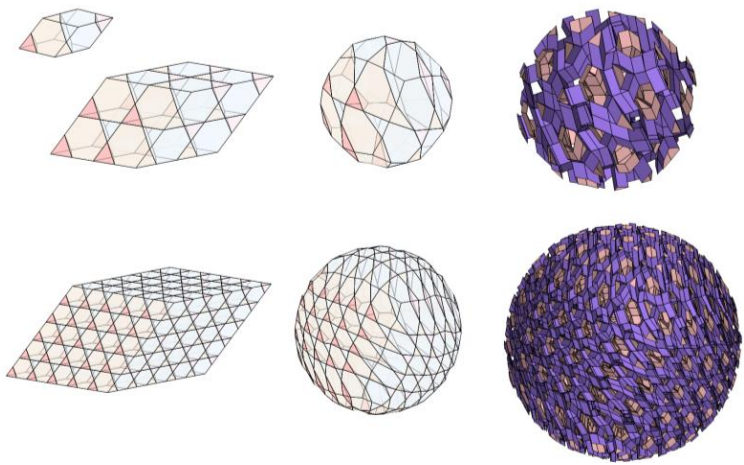

#7 octahedra and cuboctahedra (1:1)

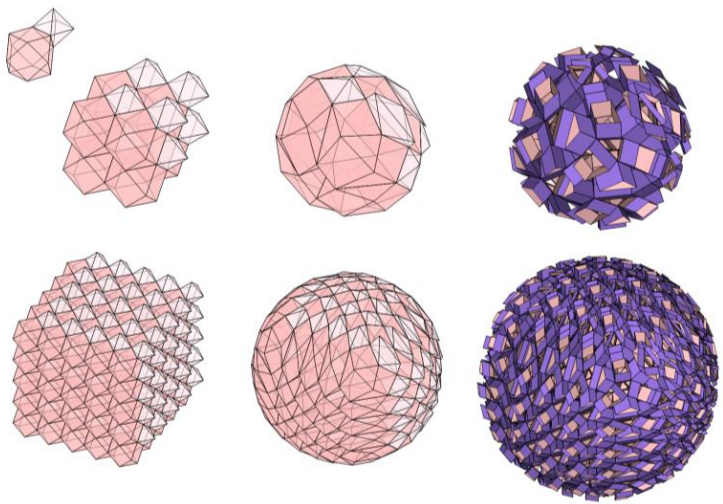

#8    octahedra and truncated cubes (1:1)

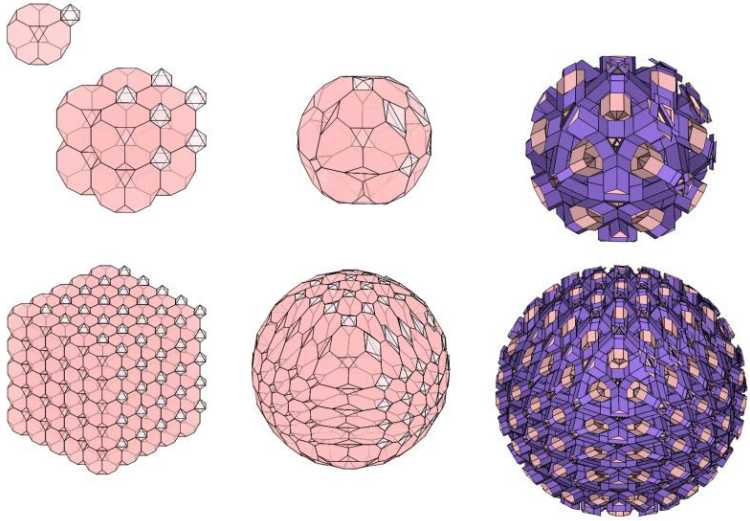

*Supplementary Fig. 5. (Continued)*

#9    cuboctahedra, rhombicuboctahedra and cubes (1:1:3)

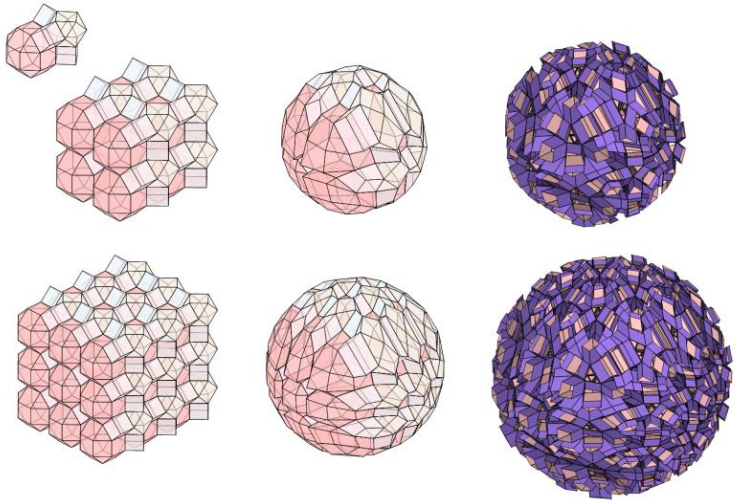

#10 cuboctahedra, truncated tetrahedra and truncated octahedra (1:1:2)

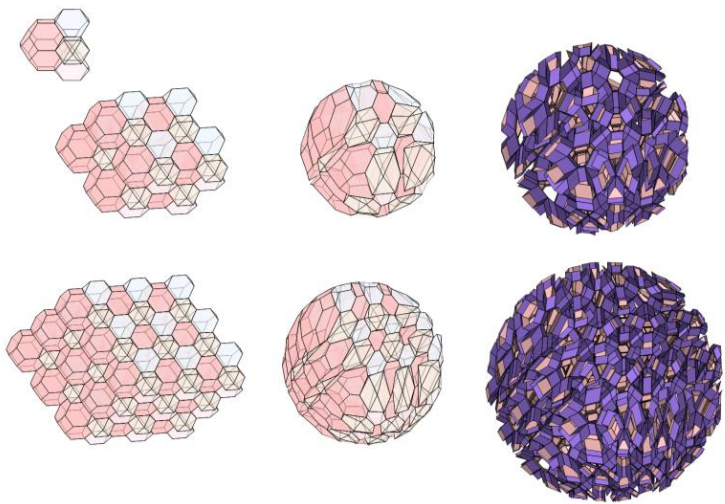

#11 Triangular prisms (2 of them comprise a unit)

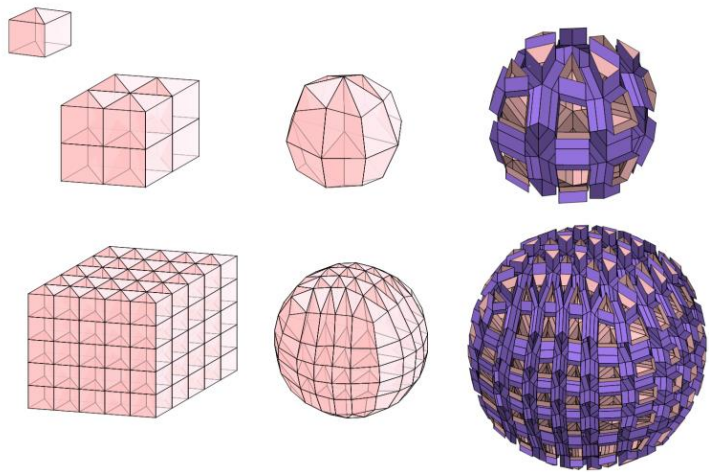

*Supplementary Fig. 5. (Continued)*

# 12  Triangular prisms (4 comprise a unit)

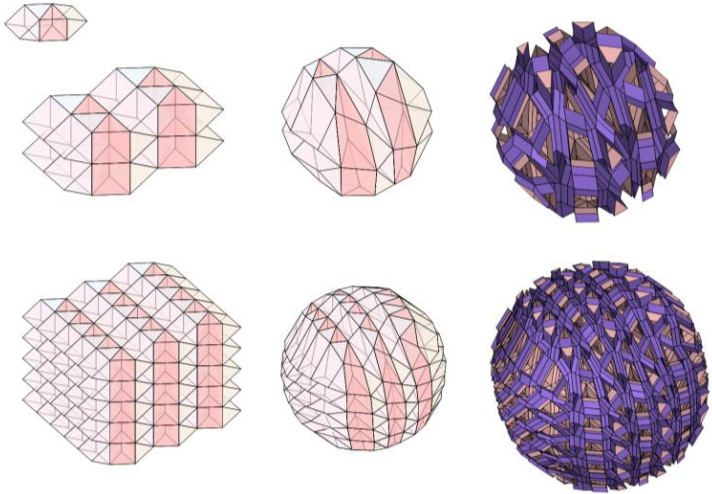

#13   triangular prisms and cubes (2:1)

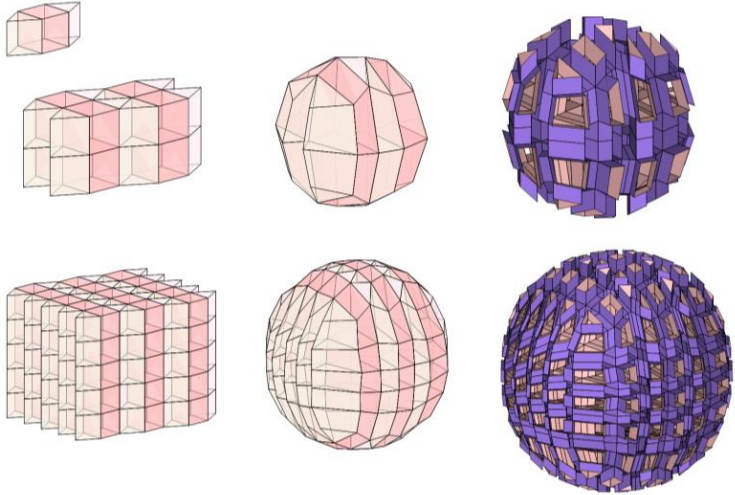

#14   triangular prisms and cubes (2:1)

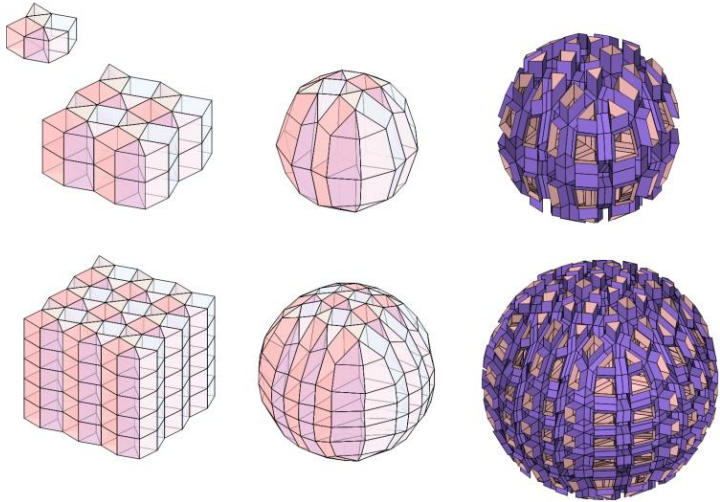

*Supplementary Fig. 5. (Continued)*

#15    triangular prisms and cubes (4:2)

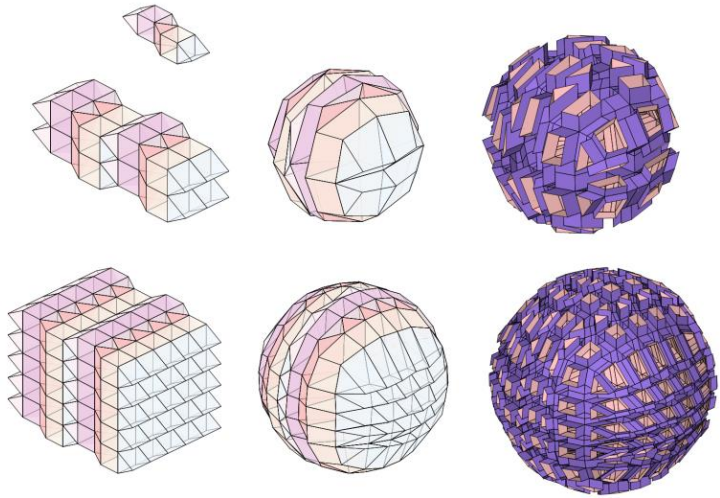

#16    triangular prims, cubes and hexagonal prisms (2:3:1)

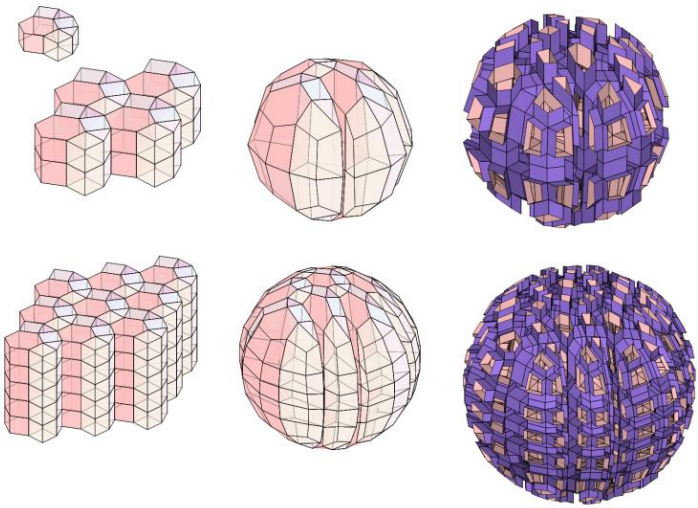

#17    triangular prisms and hexagonal prisms (8:1)

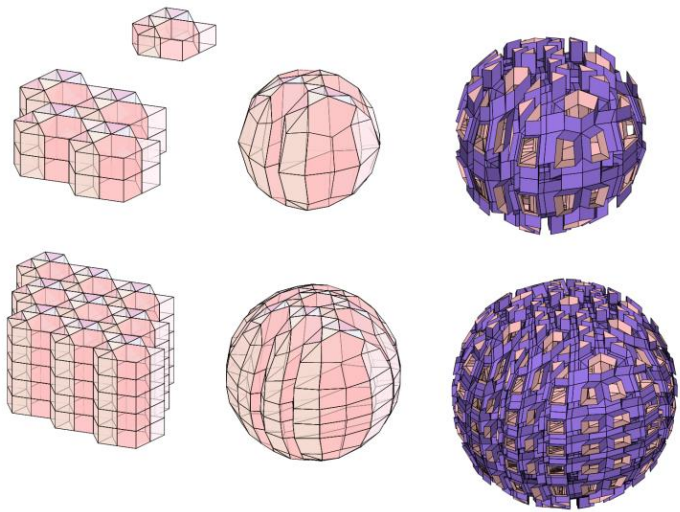

*Supplementary Fig. 5. (Continued)*

#18   triangular prisms and hexagonal prisms (2:1)

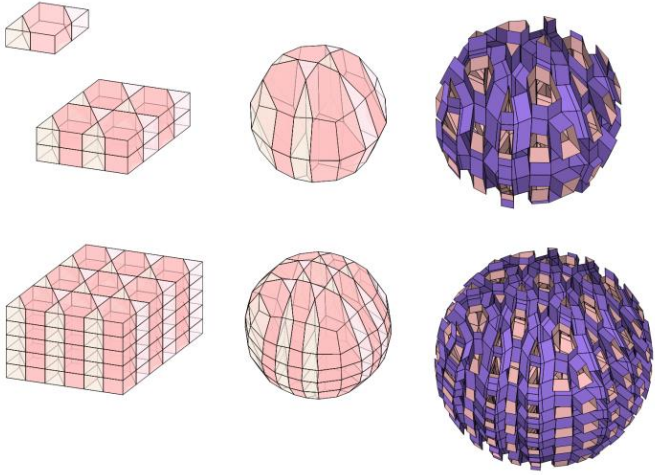

#19   triangular prisms and dodecagonal prisms (2:1)

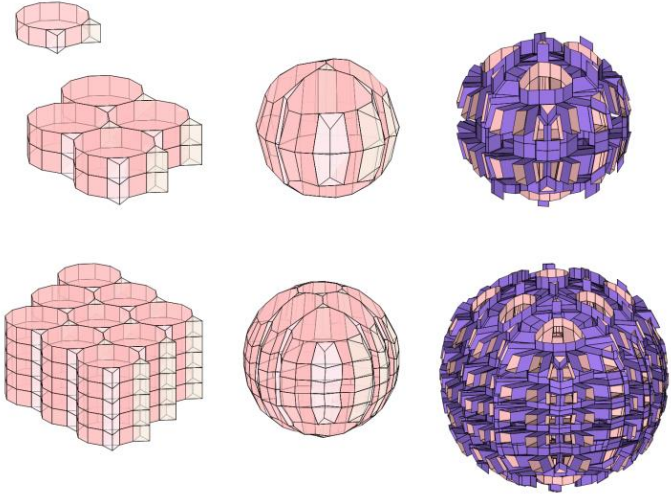

#20   rhombicuboctahedra, truncated cubes, cubes and octagonal prisms (1:1:3:3)

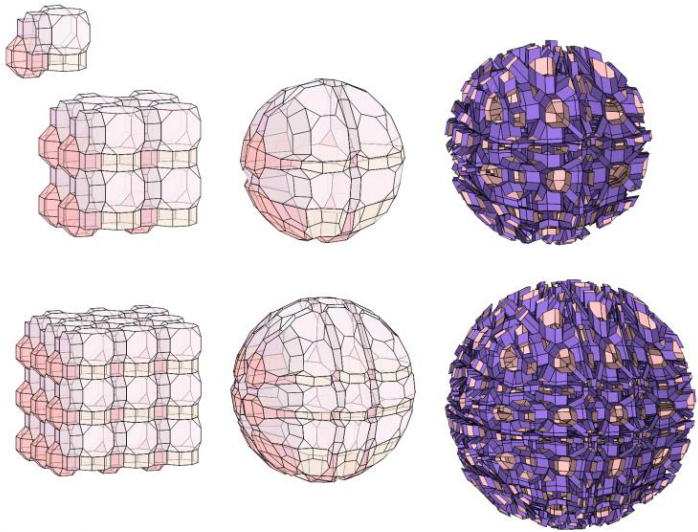

*Supplementary Fig. 5. (Continued)*

#21  truncated tetrahedra, truncated cubes and truncated cuboctahedra (2:1:1)

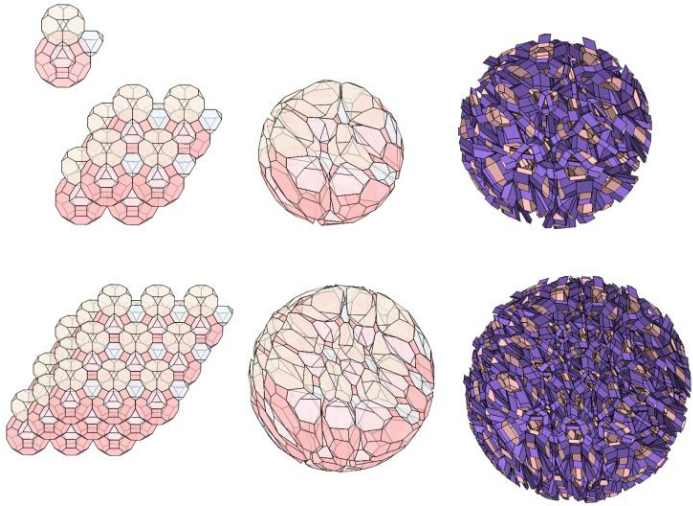

#22 cubes

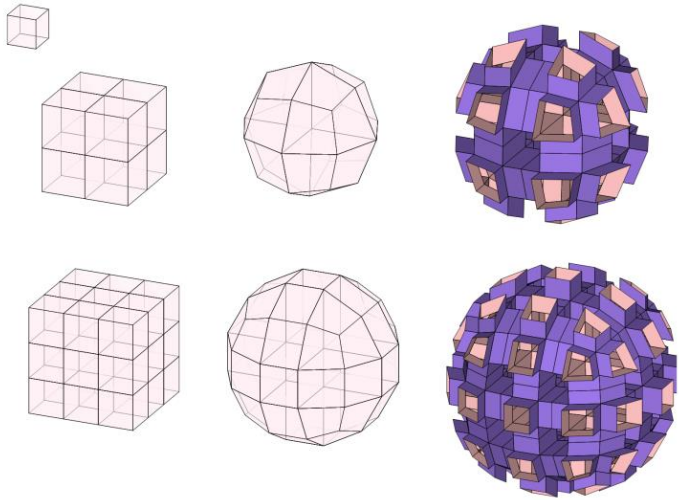

#23  cubes, hexagonal prisms and dodecagonal prisms (3:2:1)

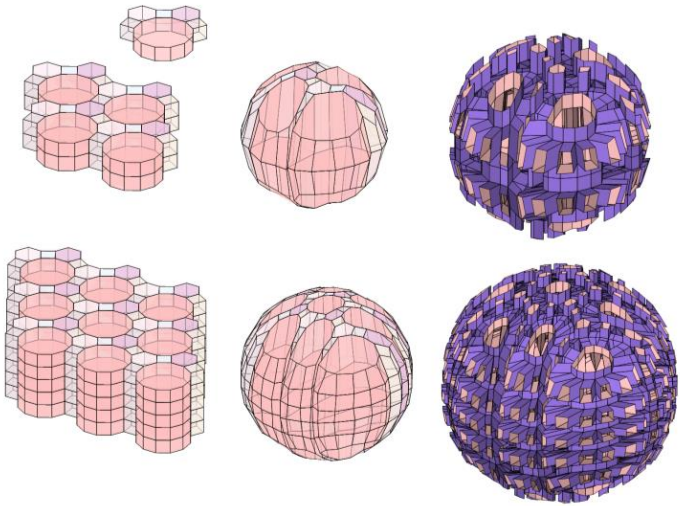

*Supplementary Fig. 5. (Continued)*

#24  cubes and hexagonal prisms (1:1)

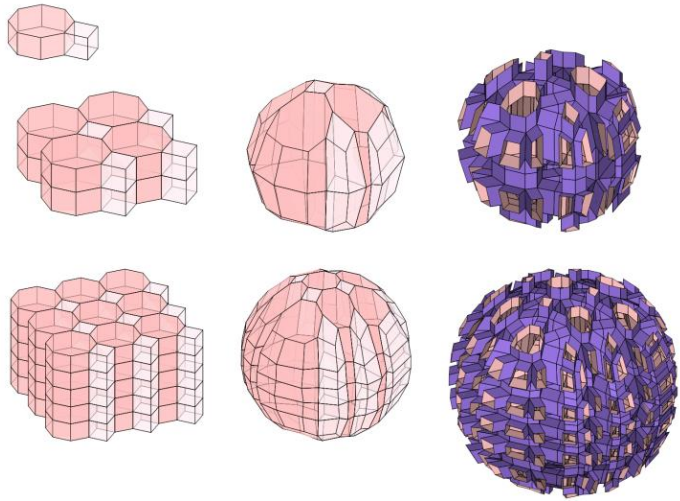

#25  cubes, truncated octahedra and truncated cuboctahedra (3:1:1)

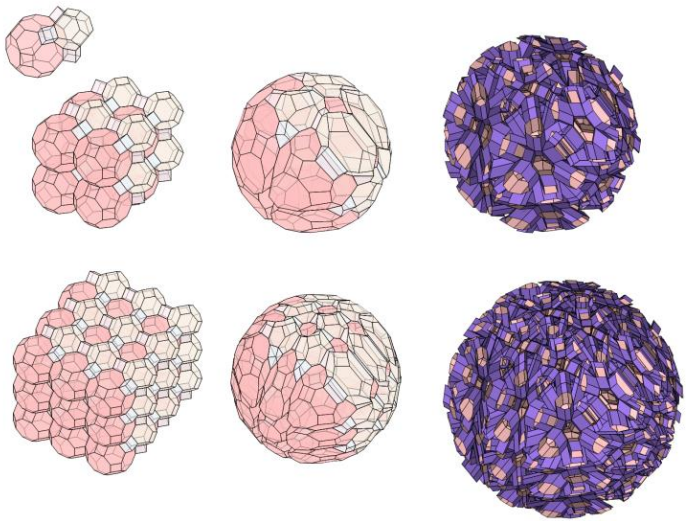

#26  hexagonal prisms

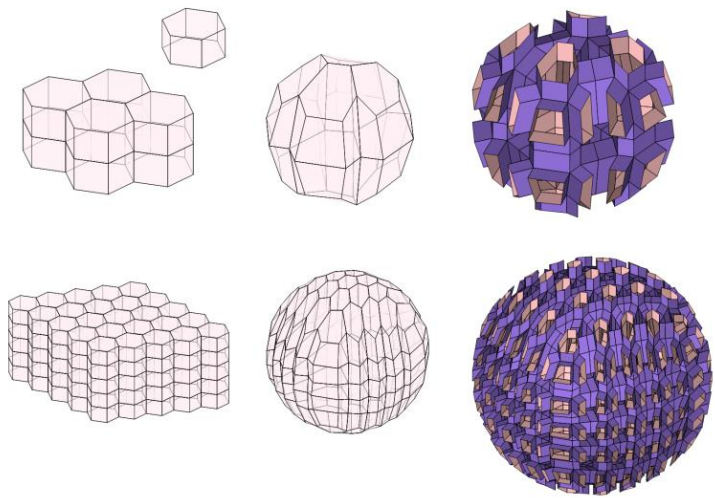

*Supplementary Fig. 5. (Continued)*

#27  octagonal prisms and truncated cuboctahedra (3:1)

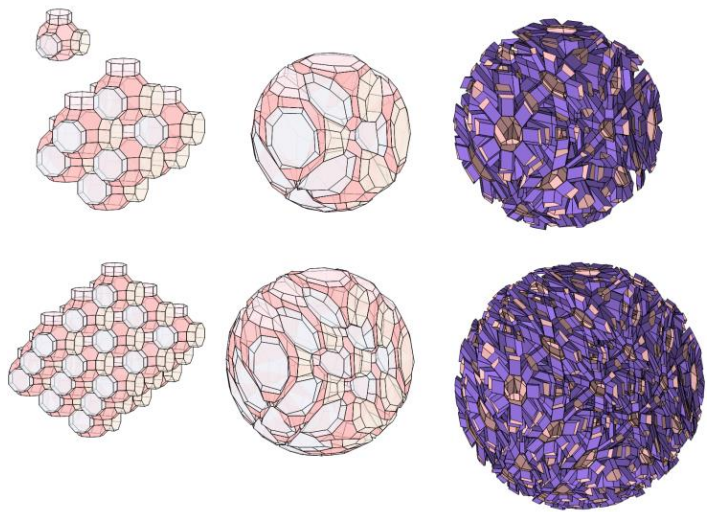

#28  truncated octahedra

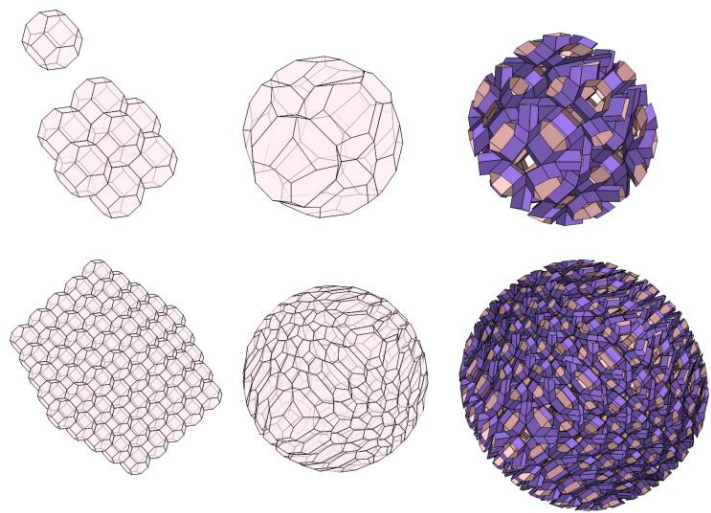

*Supplementary Fig. 5. (Continued)*